\documentclass{article}
\usepackage{graphicx} 
\usepackage{authblk}
\title{Subsampling of avalanches in the fiber bundle models of fracture}
\author{Narendra Kumar Bodaballa, Soumyajyoti Biswas }
\affil{Department of Physics, SRM University - AP, Amaravati 522240, Andhra Pradesh, India}

\begin{document}

\maketitle

\begin{abstract}
    We study the subsampling of the avalanches in the fiber bundle model of fracture. In cases where only a part of the system is observed for the micro-failure events, the recorded avalanche statistics gets distorted compared to the actual fracture events. We show that, particularly in the cases where the load redistribution is localized, this distortion is significant. Surprisingly, however, near an elastic failure regime, the distortion is minimized, suggesting a much reduced observational capacity could still represent the actual failure dynamics in the case of fracture of elastic solids. 
\end{abstract}


\section{Introduction}
The process of avalanche dynamics with scale-free size distributions, particularly for fracture, is a delicate interplay between disorders present in the material and damage localization due to stress concentration\cite{biswas2015nucleation,roy2017modes} . These two phenomena have opposing effects, when it comes to spatial spreading and temporal intermittency of the avalanches. In an ideal pure solid, according to Griffiths' theory, any existing crack will grow, provided that is energetically favorable, and will break the solid without having any other spatially non-connected damage\cite{griffith1921vi,Lawn_1993}. The other extreme is when a solid contains high range of disorder, measured through local strengths, such that the effect of damage localization due to stress concentration near a crack tip is completely masked by the fluctuations in local strengths. In the former case, the solid fails through nucleation and the latter model is best suited for a percolation picture\cite{moreira2012fracturing,shekhawat2013damage}

Most real solids, however, fall in between these two extremes. There is a spatial range that gets most affected through stress localization near an already damaged region, and then there is disorder that in effect causes further localization in damage. The result is a nominally uncorrelated microfractures that gradually forms a localized damage near the failure point. Depending on the system size and range of stress field modification near a damaged region, the localization becomes ultimately detectable (leading to eventual nucleation driven failure) or stays largely scattered damages (leading to percolation driven failure). 

Both these limiting cases, and the intermediate behavior, are well captured through simple models, such as the random fuse model and the fiber bundle model. While the random fuse model is a more realistic picture in terms of the similarity between the deformation in the electric field there and that of the stress field in solids, the fiber bundle model offers a greater flexibility in terms of tuning the range of the stress field relaxation within the model. 

Apart from the competition between the nucleation and percolation driven failures in the sample that are reflected in the resulting avalanche statistics, there is also the question of detecting all the microfracturing events in the failure dynamics. In the efforts to record the events, typically by detecting the acoustic emissions through some transducers, for multiple reasons some of the events may get missed. What are recorded then is a sub-sample of the actual events occurring in the sample. Such phenomena are well studied in other avalanching systems, such as neuronal avalanches in the brain \cite{levina2017subsampling}. Depending on the system, the sub-sampling scaling of the avalanche statistics, being distinct from the finite size scaling, is an important effect to record, as most realistic systems would have such effects in them due to partial detection of the avalanches. 
Since it is the avalanche time series that offers the most non-invasive pathways in understanding the samples proximity to catastrophic failures, the effect of sub-sampling in such time series is important to note.
Before proceeding further, we should also note here that although we have so far used avalanches as a proxy to microfracturing events, what are actually detected are the resulting energy bursts\cite{salje2014crackling}. There are situations where those two are equivalent in terms of their size distributions, but not always. 

Here we study the effect of partial sampling or sub-sampling of the avalanches in a fiber bundle model with a tunable parameter that determines the range of stress field modification. 
This enables us to determine the effect of the nucleation versus percolation interplay on sub-sampling of avalanches and thereby anticipate the most accurate representation of the actual avalanche time series from a partially observed one. 

\section{Model and methods}
\subsection{Long range load sharing model}
In studying the sub-sampling of avalanches we use a version of the fiber bundle model, which is a paradigmatic model for reproducing avalanche dynamics, at least qualitatively, in fracture of disordered solids. Particularly, we use the $\gamma$-model, where following a micro-fracture, the stress field proportional to $1/|r_i-r_j|^{\gamma}$, where $r_i$ is the location of the microfracture and $r_j$ is the location at which the stress field modification is being calculated. 

\begin{figure}[tbh]
\includegraphics[width=12cm]{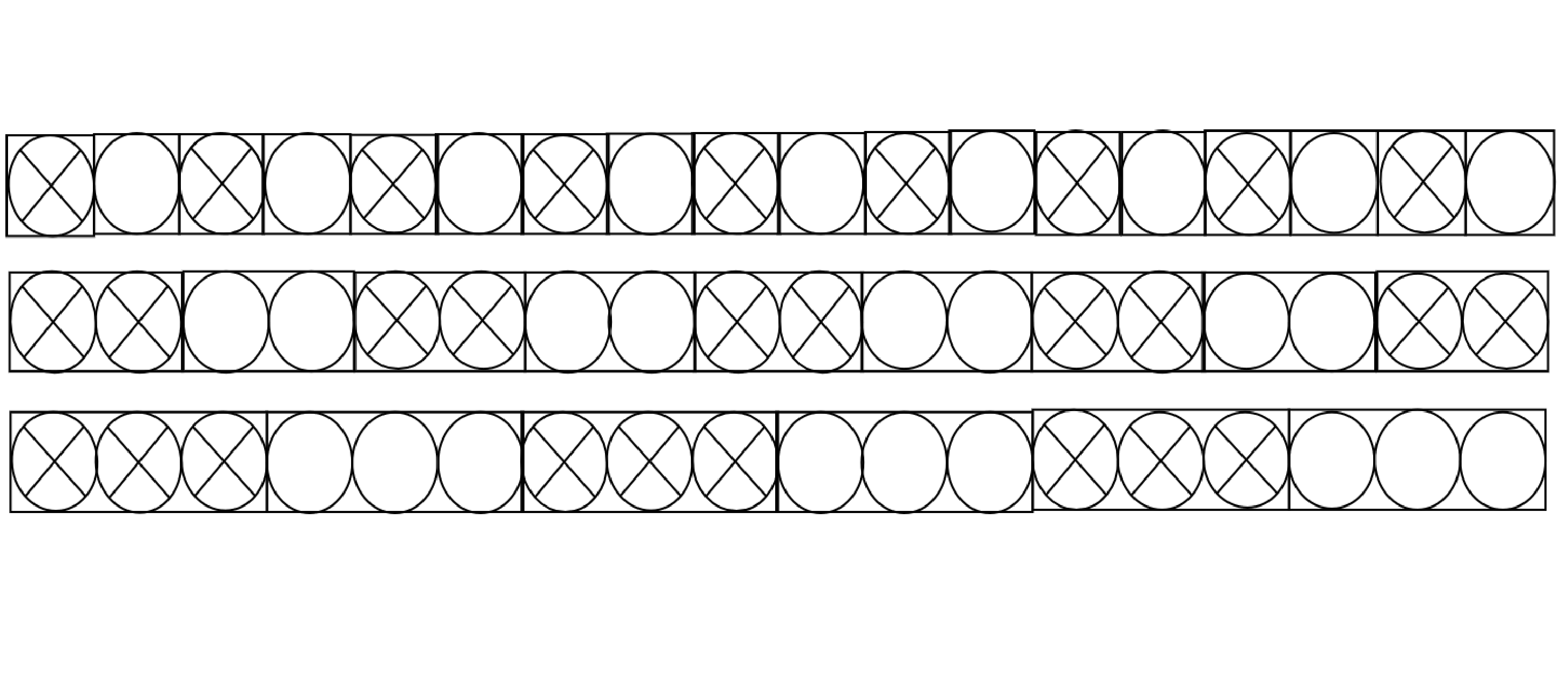}
\caption{The schematic diagram for the partially observed fiber bundle model is shown. The open circles represent the observed fibers, whereas the crossed circles represent unobserved fibers. The three cases are for $M=1$, $M=2$ and $M=3$ from the top to the bottom. In each case, the observed part is $50\%$, but the effects of having consecutive observation points are seen in the recorded avalanche statistics.}
\label{schematic}
\end{figure}

The dynamics go as follows: A set of $N$ parallel fibers are arranged in one dimension. Each fiber is linear elastic, but has a finite threshold for load carrying capacity, beyond which it fails irreversibly. The load carried by that fiber then gets redistributed among the surviving fibers according to their distance from the failed fiber following $\delta \sigma_j \propto 1/|r_i-r_j|^{\gamma}$, where $r_i$ and $r_j$ are respectively the locations of the failed fiber and the intact fiber. This can lead to further failures and the stress redistribution then again happens using the same rule. This process continues until no further breaking happens. The load on the system then again increased slowly until the next breaking event and so on. The number of fibers breaking between two stable states of the model is called an avalanche. This process continues until the entire system collapses. This is the so called quasi-static loading mechanism, which essentially assumes that there is a wide separation of time scales of internal stress relaxation and external loading. The exponent value $\gamma$ determines the range of the stress field that gets affected following a failure, and the disorder in the system comes from the failure thresholds of the individual fibers, which is drawn from a probability distribution. The width of the probability distribution sets the strength of the disorder, and for very wide distribution, the disorder strength is high and the failure dynamics crosses over from nucleation driven to percolation driven even when $\sigma$ is small. A small value for $\sigma$ obviously attempts to generated a spatially distributed damage ($\sigma=0$ being the mean field limit) and a large value for $\sigma$ generally attempts to keep the damage localized, provided the disorder strength is not too high. Such interplay for fiber bundle model and random fuse model are discussed elsewhere\cite{alava2006statistical,kawamura2012statistical} . For our purposes, we keep the disorder strength moderate by choosing a uniform probability distribution between $(0,1)$ for the failure thresholds.

The resulting avalanche dynamics is well studied. However, here we look at the effect of sub-sampling for the avalanche time series, for different values of $\gamma$. For that we choose to sample always $50\%$ of the system for recording the avalanche. That too, of course could be done in many different ways. We make a systematic study by observing $M$ consecutive fibers and then not observing $M$ consecutive fibers. This structure is repeated throughout the lattice. 

\subsection{Sub-avalanche and information measure}
Now, the meaning of observing and not observing a fiber is that when an observed fiber breaks, it is counted in an avalanche but when an unobserved fiber breaks it is not counted. Therefore, the the resulting time series of the avalanche will strongly depend on $M$ and also on $\gamma$. To quantify this difference, 
we record both the avalanches time series, one where all sites are observed and the other where part of the system is observed. We then compare the two time series. Particularly, we measure the Normalized Mutual Information (NMI) of the two series, which is a quantity that tells how much of one series could be predicted by knowing the other. Since this a normalized quantity (a complete correlation will give the value 1 and not correlated series will have the value 0), it is convenient to compare between different regions of the parameter space. 

The Normalized Mutual Information (NMI) \cite{cover1999elements,danon2005comparing}is defined as
\begin{equation}
    NMI(Y,C)=\frac{2I(Y;C)}{(H(Y)+H(C))},
    \label{nmi_def}
\end{equation}
where $I(Y;C)=H(Y)-H(Y|C)$ denotes the mutual information between $Y$ and $C$ and $H=-\sum\limits_ip_i\log_2p_i$ is the entropy and $p_i$'s are the probabilities of the individual events (avalanches), $Y$ and $C$ are the time series of the fully observed and the partially observed model. The definition is symmetric with respect to an interchange of $Y$ and $C$ though. When the complete information regarding one series can be extracted from the other, the value will be 1, and when no information can be extracted then it is zero. 
We then go on to calculate this quantity for different system size ($L$), and different $M$ values, as a function of $\gamma$.

\section{Results}
As mentioned above, we look for the quantification of the amount of information that can be extracted by partially observing the local failures in a system, which is in principle always the case due to various limitations of the measuring instruments. We quantify the correlation through Eq. (\ref{nmi_def}).

Of course, such measurements are only relevant when the stress field modification or load redistribution following the failure of a fiber is somewhat localized near the damage. For a completely mean field picture, such exercise will not matter, since there is no notion of distance.  

\subsection{Effect of spatial observation window size}
\begin{figure}[tbh]
\includegraphics[width=12cm]{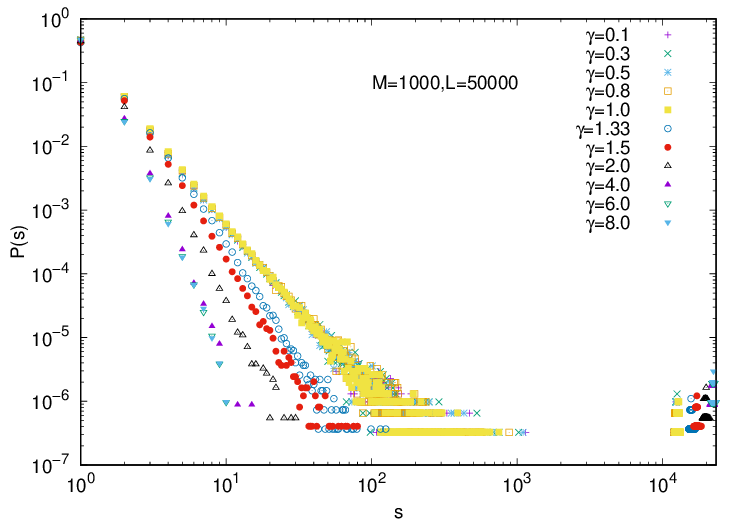}
\caption{The size distributions of avalanches are shown for various values of $\gamma$, when the parameter controlling the range of load distribution, $\gamma$, is varied. The observation patch length $M=1000$. The size distribution starts deviating from power law for $\gamma>\gamma_c$.}
\label{avl_size}
\end{figure}

First we look at the avalanche size distribution statistics when the system is partially observed with $M=1000$ and $L=50000$ (see Fig. \ref{avl_size}) for different stress localization, controlled by varying $\gamma$. For low values of $\gamma$, the system behaves as mean field, and the distribution of the avalanches, even though partially observed, show the usual $S^{-5/2}$ scaling. When $\gamma$ is increased beyond a limit, the distribution deviates from power law and becomes exponential. The cross-over to an exponential size distribution happens for $\gamma>\gamma_c\approx 1.33$, which is consistent with the behavior reported for fully observed series, where it is argued that $\gamma_c=4/3$ . For relatively large value of $M$, this is expected, since the localization of damage, when it appears, are likely to be either not captured at all (when it appears within an unobserved patch), or fully captured (when it appears within an observed patch). Therefore, the recorded avalanches are more or less similar to the case when the system is fully observed.    
\begin{figure}[tbh]
\includegraphics[width=12cm]{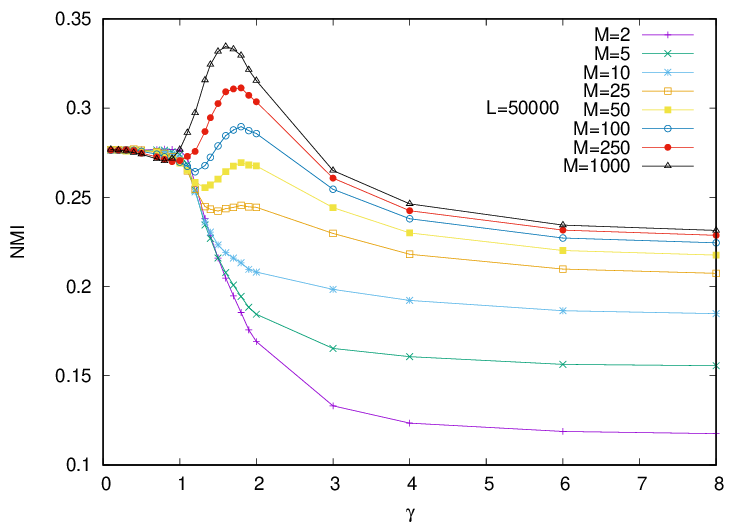}
\caption{The normalized mutual information (Eq. (\ref{nmi_def})) between the fully observed and partially observed avalanche series is plotted aginst $\gamma$ for different lengths of observation patches $M$ for a given system size $L=50000$. The mutual information show a monotonic dependence with $M$ and a non-monotonic dependence with $\gamma$. The peak appears near $-\gamma_c$, where the model is known to crossover from mean-field to local load sharing behavior. We argue that in general having a connected observation patch is more useful than not scattered observation points.}
\label{gamma_nmi}
\end{figure}
However, when $M$ values are small, a localized crack growth is unlikely to be fully within an observed patch. Therefore, there will be growing difference between the time series of the fully and partially observed samples. 

In Fig. \ref{gamma_nmi}, we show systematically the variation of the normalized mutual information between the fully and partially observed series, when $\gamma$ is varied for different $M$ values, for a given system size $L=50000$. Firstly, for low values of $\gamma$ and until $\gamma\approx \gamma_c=1.33$, the plots for different $M$ values are almost superposing. This is because, as explained above, in the mean field limit, there is no damage nucleation. Therefore, the mutual information does not depend on which part of the sample is unobserved. The values is much smaller then 1, of course, since the avalanche series of the partially observed sample is effectively obtained by randomly deleting some events and that effectively deletes some of the information. Due to the lack of spatial correlation, which is yet to develop in the model, it does not matter which of the events get deleted. On the other extreme, however, when $\gamma$ is large, the value of the mutual information increases monotonically with $M$. This is because, in that limit, the damage is almost always localized. Hence, for small $M$, it is guaranteed that the signals (avalanches) from the damaged patches are never fully captured. When $M$ is large, it becomes more and more likely that a damaged patch (or avalanches due to it) gets either fully captured or fully missed. This again leads to the same limit as the mean field, so far as the normalized mutual information is concerned.  Therefore, continuous patches of observations are more useful than scattered observations, given that the total observed fraction is fixed.
A non-monotonic behavior of the normalized mutual information is observed near $\gamma\approx 1.5$ for large $M$. This is likely due to the fact that even the damages that initiated in the unobserved part, are not fully confined within it and part of it falls within the observed part.  
\begin{figure}[tbh]
\includegraphics[width=12cm]{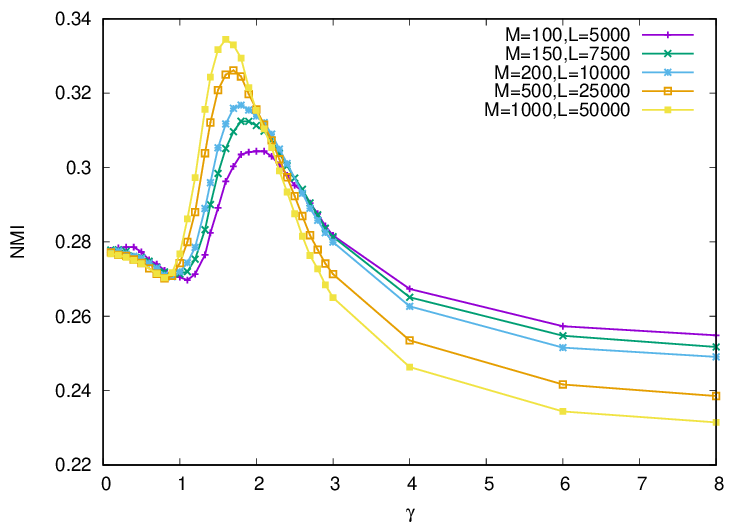}
\caption{The finite size analysis of the variation of normalized mutual information with $\gamma$, keeping the $M/L$ ratio fixed. The peak gradually shifts towards $\gamma_c=4/3$, although it starts from a higher $\gamma$ value for smaller system sizes.}
\label{gamma_nmi_L}
\end{figure}

\subsection{Effect of system size}
Finally, we look at the system size effect on the non-monotonic behavior of the mutual information between the observed and unobserved series, by keeping $M/L$ constant (see Fig. \ref{gamma_nmi_L}). This shows that the peak of the curve gradually shifts towards $\gamma_c$. We conjecture that in the large system size limit, the peak will reside at the crossover point. 

\begin{figure*}[ht]
\centering
\includegraphics[width=13cm, keepaspectratio]{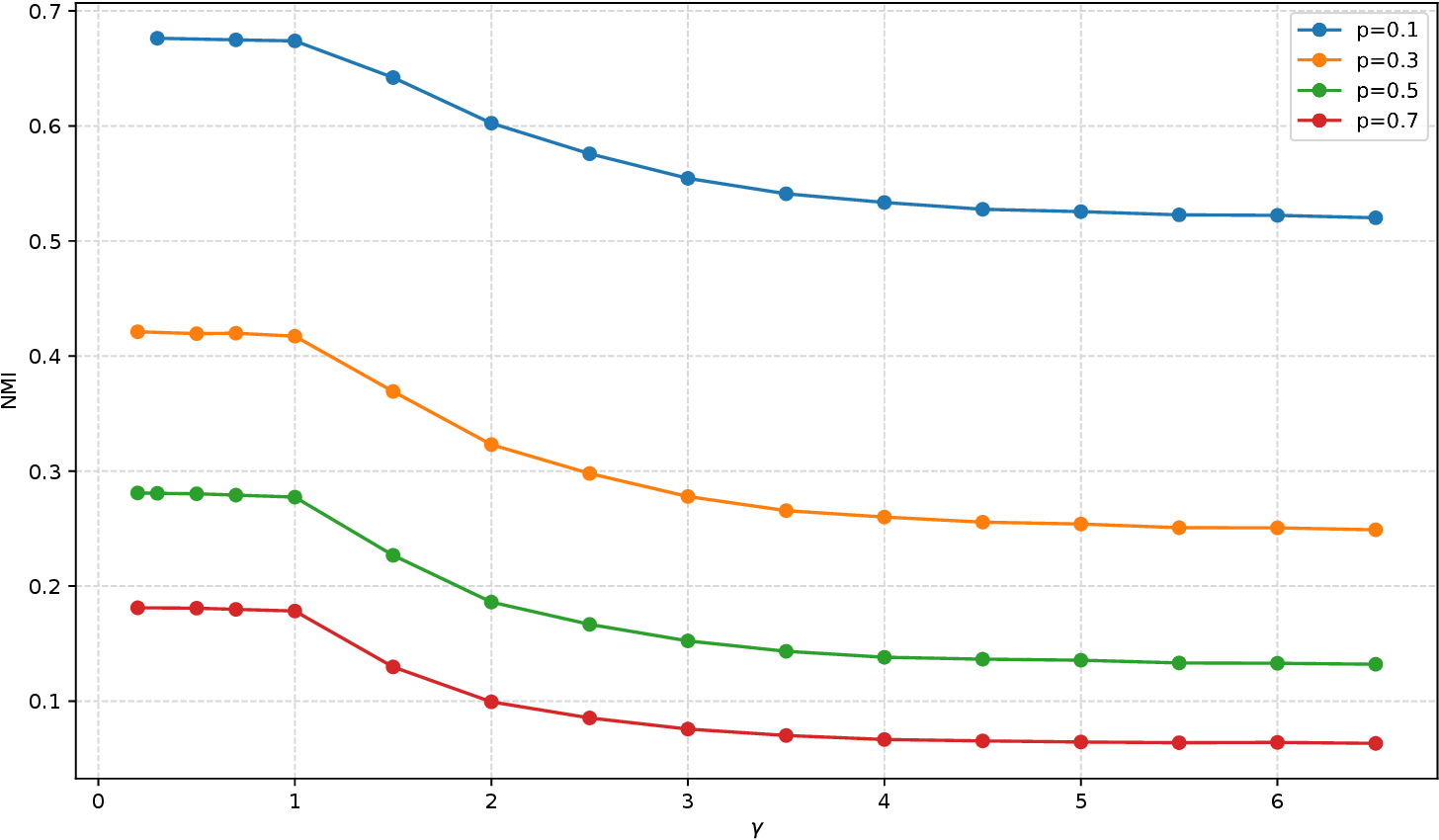}  
\caption{The variations of normalized mutual information are shown with different values of random observation fraction $p$ of the model. In all cases, there is significant drop in the NMI values beyond $\gamma=1$.}
\label{nmi_in_p}
\end{figure*}

Similarly, for a randomly sampled ($p$ fraction) system, the normalized mutual information value shows (see Fig. \ref{nmi_in_p}) significant drop beyond $\gamma=1$.

\section{Discussions and conclusion}
In failure properties of disordered solids, strength of disorder and damage nucleation play crucial roles in determining their failure statistics. In models, their interplay is often manifested in the finite size effects related to the damage statistics, leading to emergence of system-spanning correlation through percolative or nucleation of damage. Accordingly the assessments of impending hazard (catastrophic failure) depend on how the eventual failure occurs. In experimental scenarios, however, an additional consideration would be the completeness of detection of the signals (primarily as acoustic emissions) emitted from micro-damages leading up to the eventual failure. 
Partial detections, refereed to here as sub-sampling, has been an important question in characterization of statistical properties in avalanching system, and it is natural to expect such effects in the case of fracture.

The fiber bundle model with variable stress redistribution range serves as a useful testbed for checking the effects of partial observation of samples and its effect on damage statistics. We consider a regular arrangement of `detectors' in the sense that failure of some of the failed fibers are recorded within an avalanche, and the others are not. We have kept the total observed fraction at $50\%$, but the arrangements of observed fibers within the array of fibers vary (see Fig. \ref{schematic}). As a function of the load redistribution range, the model crosses over from a mean-field ($\gamma\to 0$) to a local load sharing limit ($\gamma\to\infty$) with the crossover near $\gamma_c\approx 4/3$ \cite{biswas2015nucleation,bodaballa2022correlation}. The effect of this crossover is visible in the mutual information between the fully observed avalanche series and the partially observed one. Of course, when the (normalized) mutual information is higher, the partially observed series is a more faithful representation of the fully observed series, and hence is likely to reproduce the early warning signals\cite{pradhan2005precursors,clements2019early,southall2021early}  that one can derive from such time series. Surprisingly, the normalized mutual information shows a peak near the crossover value of $\gamma$, which becomes more evident for larger system sizes. Furthermore, it is also seen that a longer patch length of observed fibers is more helpful in reproducing the original series than more scattered observations. We argue that more localized events are then completely enclosed within the observed patch than when the observation points are scattered. The competing effect, of course, is then the unobserved patches might also contain damaged regions. However, this symmetry seems to be broken near the crossover point and the normalized mutual information shows a peak. 

In conclusion, subsampling of fracture avalanches have profound effect on the resulting avalanche statistics. The connectivity of observed patches are more beneficial than having spatially scattered observation points, keeping the total observation fraction constant. 

 \section*{Acknowledgments}
 S.B. acknowledges support of SEED grant SRMAP/URG/SEED/2023-24/038. 
 
\bibliographystyle{unsrt}
\bibliography{bib}

\begin{thebibliography}{10}

\bibitem{biswas2015nucleation}
Soumyajyoti Biswas, Subhadeep Roy, and Purusattam Ray.
\newblock Nucleation versus percolation: Scaling criterion for failure in disordered solids.
\newblock {\em Physical Review E}, 91(5):050105, 2015.

\bibitem{roy2017modes}
Subhadeep Roy, Soumyajyoti Biswas, and Purusattam Ray.
\newblock Modes of failure in disordered solids.
\newblock {\em Physical Review E}, 96(6):063003, 2017.

\bibitem{griffith1921vi}
Alan~Arnold Griffith.
\newblock Vi. the phenomena of rupture and flow in solids.
\newblock {\em Philosophical transactions of the royal society of london. Series A, containing papers of a mathematical or physical character}, 221(582-593):163--198, 1921.

\bibitem{Lawn_1993}
Brian Lawn.
\newblock {\em Fracture of Brittle Solids}.
\newblock Cambridge Solid State Science Series. Cambridge University Press, 2 edition, 1993.

\bibitem{moreira2012fracturing}
AA~Moreira, CLN Oliveira, A~Hansen, NAM Ara{\'u}jo, HJ~Herrmann, and JS~Andrade~Jr.
\newblock Fracturing highly disordered materials.
\newblock {\em Physical review letters}, 109(25):255701, 2012.

\bibitem{shekhawat2013damage}
Ashivni Shekhawat, Stefano Zapperi, and James~P Sethna.
\newblock From damage percolation to crack nucleation through finite size criticality.
\newblock {\em Physical review letters}, 110(18):185505, 2013.

\bibitem{levina2017subsampling}
Anna Levina and Viola Priesemann.
\newblock Subsampling scaling.
\newblock {\em Nature communications}, 8(1):15140, 2017.

\bibitem{salje2014crackling}
Ekhard~KH Salje and Karin~A Dahmen.
\newblock Crackling noise in disordered materials.
\newblock {\em Annu. Rev. Condens. Matter Phys.}, 5(1):233--254, 2014.

\bibitem{alava2006statistical}
Mikko~J Alava, Phani~KVV Nukala, and Stefano Zapperi.
\newblock Statistical models of fracture.
\newblock {\em Advances in Physics}, 55(3-4):349--476, 2006.

\bibitem{kawamura2012statistical}
Hikaru Kawamura, Takahiro Hatano, Naoyuki Kato, Soumyajyoti Biswas, and Bikas~K Chakrabarti.
\newblock Statistical physics of fracture, friction, and earthquakes.
\newblock {\em Reviews of Modern Physics}, 84(2):839--884, 2012.

\bibitem{cover1999elements}
Thomas~M Cover.
\newblock {\em Elements of information theory}.
\newblock John Wiley \& Sons, 1999.

\bibitem{danon2005comparing}
Leon Danon, Albert Diaz-Guilera, Jordi Duch, and Alex Arenas.
\newblock Comparing community structure identification.
\newblock {\em Journal of statistical mechanics: Theory and experiment}, 2005(09):P09008, 2005.

\bibitem{bodaballa2022correlation}
Narendra~K Bodaballa, Soumyajyoti Biswas, and Subhadeep Roy.
\newblock Correlation between avalanches and emitted energies during fracture with a variable stress release range.
\newblock {\em Frontiers in Physics}, 10:768853, 2022.

\bibitem{pradhan2005precursors}
Srutarshi Pradhan and Bikas~K Chakrabarti.
\newblock Precursors of catastrophic failures.
\newblock In {\em Nonequilibrium Phenomena in Plasmas}, pages 293--310. Springer, 2005.

\bibitem{clements2019early}
Christopher~F Clements, Michael~A McCarthy, and Julia~L Blanchard.
\newblock Early warning signals of recovery in complex systems.
\newblock {\em Nature Communications}, 10(1):1681, 2019.

\bibitem{southall2021early}
Emma Southall, Tobias~S Brett, Michael~J Tildesley, and Louise Dyson.
\newblock Early warning signals of infectious disease transitions: a review.
\newblock {\em Journal of the Royal Society Interface}, 18(182):20210555, 2021.

\end{thebibliography}

\end{document}